\documentclass[conference]{IEEEtran}

\usepackage{graphicx}
\usepackage{amsfonts}
\usepackage[bookmarks=false]{hyperref}

\begin{document}
\title{Spectral Visibility Graphs:\\Application to Similarity of Harmonic Signals }
%\title{Spectral visibility graphs: a new representation and its application to similarity of harmonic signals }

\author{\IEEEauthorblockN{Delia Fano Yela, Dan Stowell and Mark Sandler}
\IEEEauthorblockA{Centre for Digital Music\\
Queen Mary University of London\\
E14NS, London, UK\\
Email: d.fanoyela@qmul.ac.uk}
% \and
% \IEEEauthorblockN{Dan Stowell}
% \IEEEauthorblockA{Centre for Digital Music\\
% Queen Mary University of London\\
% E14NS, London, UK\\}
% \and
% \IEEEauthorblockN{Mark Sandler}
% \IEEEauthorblockA{Centre for Digital Music\\
% Queen Mary University of London\\
% E14NS, London, UK\\}
}


%Telephone: (800) 555--1212, Fax: (888) 555--1212}


\maketitle

\begin{abstract}
Graph theory is emerging as a new source of tools for time series analysis. One promising method is to transform a signal into its visibility graph, a representation which captures many interesting aspects of the signal. Here we introduce the visibility graph for audio spectra and propose a novel representation for audio analysis: the spectral visibility graph degree. Such representation inherently captures the harmonic content of the signal whilst being resilient to broadband noise. We present experiments demonstrating its utility to measure robust similarity between harmonic signals in real and synthesised audio data. The source code is available online.
\end{abstract}


% \fi
\IEEEpeerreviewmaketitle

\section{Introduction}
Graphs are a tool of growing interest in the signal processing community for data representation and analysis. Their structure offers a new perspective, often unveiling non trivial properties on the data they represent. In particular, time series analysis has greatly benefited from graph representations as they provide a mapping able to deal with non-linearities and multi-scaling issues present in multiple applications  \cite{BarberC2010_Graphicalmodelstime_ISPM,StowellP2013_Segregatingeventstreams_TJoMLR,NunezLGL12_ReviewVg_InTech,CampanharoSMRA11_dualityTS_Pone}.

A popular mapping from time series to complex networks is the visibility graph \cite{LacasaLBLN08_VisibilityGraph_PNAS}. Every node in such graph represents a datum of the time series, and two nodes are connected if they fulfil visibility criteria analogous to the visibility between points on a landscape. The visibility between data will only depend on their relative height and location, creating a graph structure capturing the links between data. The success of this simple visibility mapping is partly due to its powerful properties. Visibility graphs preserve characteristics of the time series such as periodicity \cite{NunezLVGL12_PeriodicityHVg_IJBC}, and are invariant to several transformations of the time series, such as vertical and horizontal rescaling. It was introduced as a time series analysis tool \cite{LacasaLBLN08_VisibilityGraph_PNAS, LacasaNL15_multiplex_Nature} and has been successfully employed in several applications such as financial series analysis \cite{musmeci2017multiplex}. 

Here we introduce visibility graphs applied to magnitude spectra. Such graph will preserve all the properties of visibility graphs as its construction remains the same. Therefore, similarly to time series, the visibility graph of spectra may reveal hidden structures in the signal not apparent in the magnitude domain. In particular, we focus on musical audio signals, and we propose the spectral visibility graph degree as a novel representation for audio analysis.

In  the  spectrum  of  audio  signals,  peaks often correspond  to harmonic events while percussive or burst-like events present a broadband nature.
Broadband content can be a nuisance in music analysis tasks when the target is the harmonic content of the signal. 
In particular, tasks that require distances of harmonic content face a challenge when the broadband event out-powers the targeted harmonic one \cite{FanoYelaEFS17_HybridKamNmf_ICASSP}.

Conversely, we will show that the spectral visibility graph degree has properties which preserve the harmonic peaks salience in presence of broadband events. Therefore, we propose to use such representation for audio analysis as an alternative to the commonly used spectrogram. One can think of the proposed representation as a `tidied-up' version of the spectrogram, that could find applications as an alternative starting point for harmonic similarity estimation methods such as \cite{bello2005robust} or audio fingerprinting methods based on peak-picking \cite{wang2003industrial,sonnleitner2014quad}. Here we demonstrate its use for robust similarity measures of harmonic signals. 

In the experiments section we demonstrate that conventional distance metrics fail to recognise the harmonic content in the spectrum in presence of broadband noise, whereas the proposed visibility representation is sensitive to its harmonic content. Furthermore we show how real world scenarios could benefit from such visibility representation, in a final source specific query task to retrieve vocals within a musical mixture.

\section{Visibility Graphs}

% DIAGRAM invariant to vertical rescaling --------------------------------------------------------------------------
\begin{figure*}[h]
  %\hspace*{-0.65in}
  \begin{center}
  \includegraphics[scale=0.4]{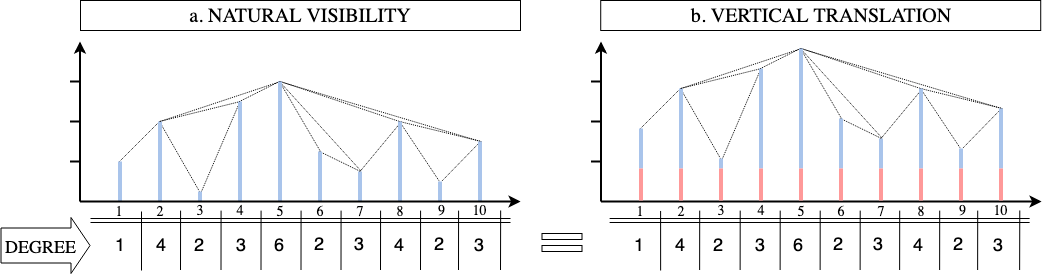}
  \end{center}
  \caption{Illustration of the visibility invariance to vertical translation. }
  \label{fig:invariant}
   \vspace{-0.2cm}
\end{figure*}
%------------------------------------------------------------------------------------------------------------------

A graph consists of a non-empty finite set of elements called nodes and a finite set of edges joining pairs of nodes together. If the set of edges is comprised of ordered pairs of distinct nodes, the graph is called a digraph and it is said to be directed. On the other hand, if the connection between nodes is symmetric, the graph is said to be undirected \cite{bang2008digraphs}. 

The visibility graph described in \cite{LacasaLBLN08_VisibilityGraph_PNAS} is associated to a time series although valid for any ordered sequence. Every datum is defined as a node in the graph and two pairs of nodes are joined by an edge if they are visible to each other. The visibility between two points $(t_{a}, y_{a})$ and $(t_{b},y_{b})$ of a given time series $y=f(t)$ of length $N$, is determined by the following geometrical criterion:
\[
y_{c} < y_{a} + ( y_{b} - y_{a} ) \frac{t_{c} - t_{a} }{t_{b} - t_{a}}
\]
where $(t_{c}, y_{c}) $ is every intermediate point such that $t_{a}<t_{c}<t_{b}$. In other words, two points of a given time series are said to `see' each other if one can draw a straight line joining them without intercepting any intermediate data height. 

This visibility is referred to as \emph{natural} visibility, as other kinds exist \cite{LuqueLBFL09_HVg_PRE}. Here we will use the defined \emph{natural} visibility and simply refer to it as \emph{visibility}. Since such visibility is symmetric (both points either see each other or do not) the visibility graph is an undirected graph. Visibility graphs are always fully connected (i.e. every node has at least one edge) as every datum always sees at least its neighbours. Note that the visibility transformation is not reversible, finding a greater utility as an analysis tool.

We can represent a visibility graph in the form of a square binary matrix $ A(i,j) \in \mathbb{B}^{N \times N}$ ($ i, j =1,2,3,... , N$ ) such that:
\[
A(i,j) =  1 \Leftrightarrow nodes \ \ i \ \ and \ \ j  \ \ are \ \ visible
\]
\[
A(i,j) =  0 \ \ otherwise
\]
This matrix $A$ is referred to as \emph{adjacency} matrix. Since the visibility graph is undirected, the corresponding adjacency matrix will be symmetric.

The degree $k(i)$ of a node $i$ is defined as the count of its edges, in other words, the number of nodes connected to it. In the case of visibility graphs, the degree of a node indicates the number of visible nodes or data points. For example, in Figure \ref{fig:invariant}, the first value of the sequence only sees its neighbour and so its degree will be equal to $1$. However, the maximum data point of the sequence in fifth position has a wider view and therefore has a larger degree value. 

The degree of a node can easily be obtained from the adjacency matrix as it corresponds to the sum of either the rows or columns (indifferent in this symmetric case) storing the edges of that node. We can define a degree vector $k \in \mathbb{N}^{N}$ containing the degrees of all nodes $i =1,2,3,... , N$ of the visibility graph with adjacency matrix $A$ as follows:
\[
k(i) = \sum_{j=1}^{N} A(i,j)
\]

We also define the degree distribution $p$, indicating how often the different degree values appear in the degree vector (i.e. histogram). If the values are normalised by the total number of nodes in the graph, $p$ will represent the probability of the different degree values of existing in that graph.

Visibility graphs are invariant to horizontal and vertical translation as the absolute value of the data points have no effect on the visibility (only their relative values matter). For instance, as illustrated in Figure \ref{fig:invariant}, the visibility of a signal with and without a DC offset is equal and so the degree value for each node remains the same in both cases.
Furthermore, rescaling of both horizontal and vertical axes also has no effect on the visibility. If the signal is time stretched, the relative position of the points remains the same and so does the visibility. 

\section{Spectral Visibility Graphs for Audio Signals}

% Spectrogram Vs spectral visibility graph degree  --------------------------------------------------------------------------
\begin{figure*}[t]
  \begin{center}	
  \includegraphics[scale=0.6]{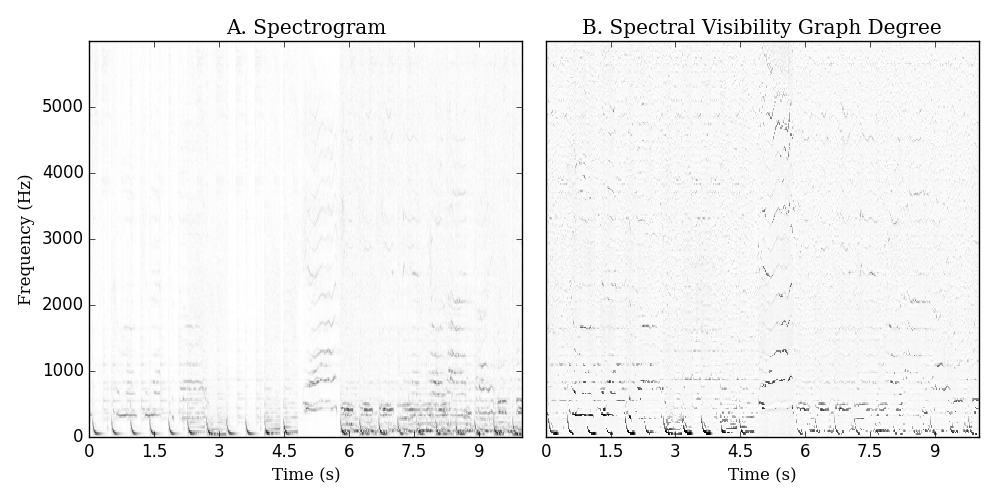}
  \end{center}
  \caption{ The spectrogram (A) and the proposed representation (B) of 10 seconds of track 51 of the dataset DSD100. Both representations are normalised by their own maximum and compressed by a factor of 0.6. The spectral visibility graph degree enhances the harmonics components of the signal.}
  \label{fig:rep}
   \vspace{-0.2cm}
\end{figure*}
%-----------------

Inspired by the invariant properties of visibility graphs, we propose to employ such mapping for magnitude spectra, introducing visibility graphs to spectral analysis. We define the spectral visibility graph (SVg) of a given magnitude spectrum $\bar{s}= f(\omega)$ of $s \in \mathbb{C}^{F}$, where $\omega$ denotes frequency and $F$ is the total number of frequency bins, following the construction of visibility graphs for time series. 

Every time-frequency bin corresponds to a node, unlike previous audio graph-based representations where nodes are associated to feature vectors or time frames  \cite{mcfee2014analyzing,dFanoYela_LVA}. Two nodes are connected if the associated frequency bins $(\omega_{a}, \bar{s}_{a})$ and $(\omega_{b}, \bar{s}_{b})$ see each other, fulfilling the visibility criterion:
\[
\bar{s}_{c} < \bar{s}_{a} + ( \bar{s}_{b} - \bar{s}_{a} ) \frac{\omega_{c} - \omega_{a} }{\omega_{b} - \omega_{a}}
\]
where $(\omega_{c}, \bar{s}_{c}) $ is every intermediate frequency bin such that $\omega_{a}<\omega_{c}<\omega_{b}$. 
Similarly to time series visibility graphs, we can analogously construct its associated adjacency matrix $A$ and find the degree $k$ and degree distribution $p$ vectors, such that for $f = 1,2,..., F$ frequency bins in the spectrum:
\[
k(f) = \sum_{j=1}^{F} A(f,j)
\]
Similarly to the degree vector of time series, the SVg degree vector remains invariant under several transformations of the spectrum, including vertical and horizontal translation as well as vertical and horizontal rescaling.

%SYNTH QUERY --------------------------------------------------------------------------
\begin{figure*}
  %\hspace*{-0.65in}
   \begin{center}	
  \includegraphics[scale=0.4]{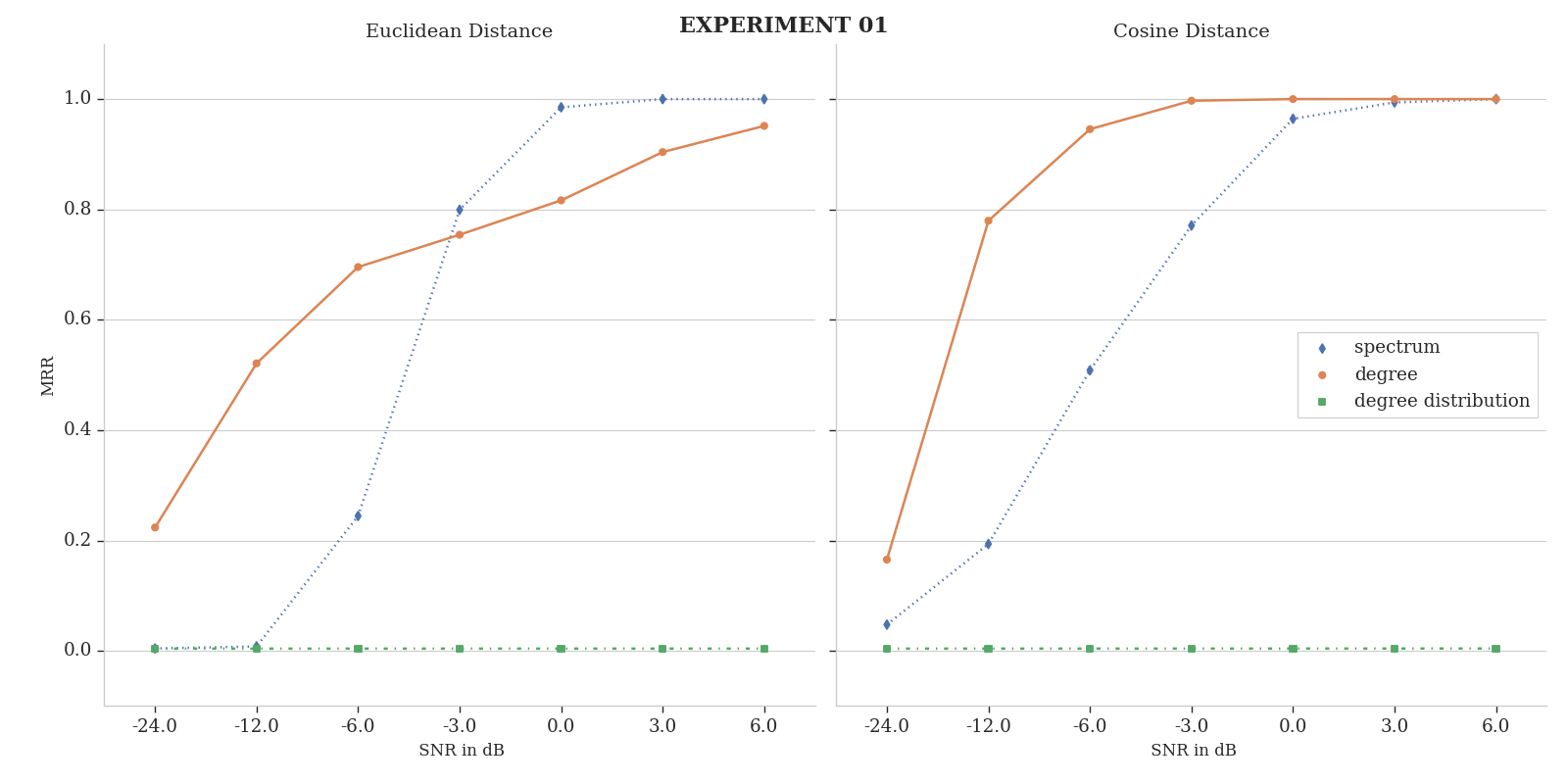}
   \end{center}	
  \caption{ The average mean-reciprocal-rank (MRR) amongst all notes of all instruments in experiment 01: 12 synthesised instruments playing 14 notes, clean and with additive random noise. Pair-wise similarity between all signals in the frequency magnitude, degree and degree distribution domain. The clean notes act as query and the expected closest neighbour is their noisy version. 
  }
  \label{fig:exp01}
  \vspace{-0.2cm}
\end{figure*}
%------------------------------------------------------------------------------------------------------------------

In the case of audio signals, a horizontal rescaling of the spectrum would correspond to a change in pitch and a vertical translation to the presence of uniform broadband noise. Being resilient to such transformations is a major advantage in the audio analysis of applications where the relation between peaks (i.e. harmonic content) is the subject of interest. Therefore, we propose the SVg degree vector $k$ as an alternative representation for magnitude spectra $\bar{s}$.

Taking a step further, let $S \in \mathbb{C}^{F \times T}$ be the spectrogram of an audio time signal $y$, and $\bar{S}$ its magnitude, where $F$ is the number of frequency bins and $T$ the number of time frames. Here, the proposed representation $K \in \mathbb{N}^{F \times T}$ will take a matrix form such that every column $t=1,2,...,T$ will correspond to the degree vector $k_{t}$ of the visibility graph of frame $t$ of $\bar{S}$ (Figure \ref{fig:rep}). More precisely, taking $A_{t} \in \mathbb{B}^{F \times F} $ as the visibility graph's adjacency matrix of the time frame's magnitude spectra $t$ (i.e. column) of the spectrogram $\bar{S}$, we define the degree matrix $K$ associated to $\bar{S}$ such that:
\[
K(f,t) = \sum_{j=1}^{F} A_{t}(f,j)
\]
where $f=1,2,..., F$ and $t = 1,2,..., T$.  We propose to use $K$ as an alternative representation to the spectrogram $S$ .

Even though spectral peaks tend to take high values in the proposed representation, their prominence will depend on their surroundings. In other words, peaks close to each other will have less height than sparse ones, such as the harmonics of a musical note. 
Looking at Figure \ref{fig:invariant}, one may notice how the height at position 4 lost pertinence in the degree domain, going from being the second maxima to being equal to lesser heights (7 and 10); explained by its proximity to the maximum peak in 5. On the other hand, the heights at position 2 and 8 (equally spaced from the maximum) surrounded by smaller heights, gained relevance in the degree domain.
Therefore, one can think the transformation into the degree domain, and so into the proposed representation, as a sort of compression enhancing sparse peaks (i.e. harmonics) visible in Figure \ref{fig:rep}. 

As an audio analysis tool, the structure and properties of the proposed mapping directly relate to harmonic content analysis, and so we propose to examine the common case where both harmonic and broadband events overlap. In such scenario, the harmonic energy in the spectrum will remain prominent up to a certain signal-to-noise ratio (SNR), taking the harmonic event as the signal of interest and the broadband as noise. If the broadband event overpowers the harmonic content, it will overcast the harmonic contribution in the magnitude spectrum, complicating the analysis of its harmonic content. 

A common task in audio analysis is the search for similar harmonic content between spectra (e.g. time frames in a spectrogram). 
In the presence of powerful additive broadband noise, most distance metrics fail to recognise the similarity of the harmonic content as they treat all spectral energy as equivalent. 
Such scenario relates to a vertical translation of the magnitude spectrum and so the harmonic event spectrum with and without additive broadband noise should present a comparable visibility graph and degree vector. Therefore, unlike in the magnitude spectrum (e.g. Figure \ref{fig:rep}.A), the harmonic peaks in the proposed representation (e.g. Figure \ref{fig:rep}.B) will remain salient in presence of additive broadband events, and so, one can now use standard distance metrics (e.g. l1 or l2norm) to reliably measure harmonic similarity. Hence we propose the SVg degree as a novel domain for robust harmonic similarity measure in audio signals. 

\section{Experiments}

To evaluate the proposed representation of audio signals for harmonic similarity measure we performed two experiments, one with synthesised data and a second one with real musical recordings. In both experiments the task is to find the correct nearest neighbour of a given harmonic event. We use three different representations of the audio signals: the magnitude spectrum, the SVg degree and the SVg degree distribution. 
Our proposed representation is the SVg degree; however, we included the degree distribution in the experiments as it has an additional pitch invariance that could benefit the task (i.e. the absolute location of peaks information is ignored). Since our goal is to compare these representations, we employ simple distance metrics (i.e. Euclidean and cosine) to conclude on which representation is more appropriate for harmonic similarity measurements. We use the mean reciprocal rank (MRR) as the evaluation metric, as we know before hand which is the one and only correct nearest neighbour.

For high frequency resolution spectra, the basic computation of visibility graphs \footnote{Original visibility graphs Fortran 90/95 implementation can be found at \url{http://www.maths.qmul.ac.uk/~lacasa/Software.html}} is not ideal ($O(n^2)$). Therefore, here we used a significantly faster alternative to compute visibility graphs based on a `Divide \& Conquer' approach ($O(n\log n)$) \cite{Lan15_DC_Caos}. Python source code for our implementation and experiments is freely available online \footnote{Available at \url{https://github.com/delialia/vgspectra}}.

In the first experiment we used part of the synthesised data from \cite{FanoYelaEOS18_ShiftInvariant_ICASSP}: 12 synthesised instruments with the same midi file of 14 notes (A2 to G4) sampled at 44100Hz. Each instrument signal was divided into the distinct midi notes and then individually transformed into the magnitude frequency domain with a Fourier transform of size 16384, `clean' spectra. Only the first 2000 bins of the magnitude spectra were kept for the rest of the analysis. 

Random normal noise was then added to the note signals at different SNR values and the result transformed to the frequency domain, `noisy' spectra. The pair-wise distances between all spectra, both clean and noisy, were then computed and sorted in ascending order. For every clean track, the rank of its noisy version was found and used to compute the MRR. This procedure is repeated for the spectral visibility graph degree representation as well as for the degree distribution. 

The average MRR across all notes of all instruments for different SNR is plotted in Figure \ref{fig:exp01}. As expected the proposed method (orange solid line) achieves best results when the SNR is low. However we see a small dip in performance relative to the raw spectrum, using the Euclidean distance in the higher SNR cases. This can be explained by the bigger difference in value between the degree peaks of the clean and noisy signals than in the spectrum case. Even though the peaks remain prominent in the noisy case, the number of nodes the `peak node' sees is reduced compared to the clean peak degree as there are new data heights induced by the noise. In the case of high SNR, the noise does not overpower the harmonic content and so it does not introduce too much of a difference in the Euclidean distance. However, the location of the peaks are better preserved in the proposed representation and so it always presents the best results whilst using the cosine distance metric.

%VOCAL QUERY --------------------------------------------------------------------------
\begin{figure}[]
	\begin{center}
		\includegraphics[scale=0.35]{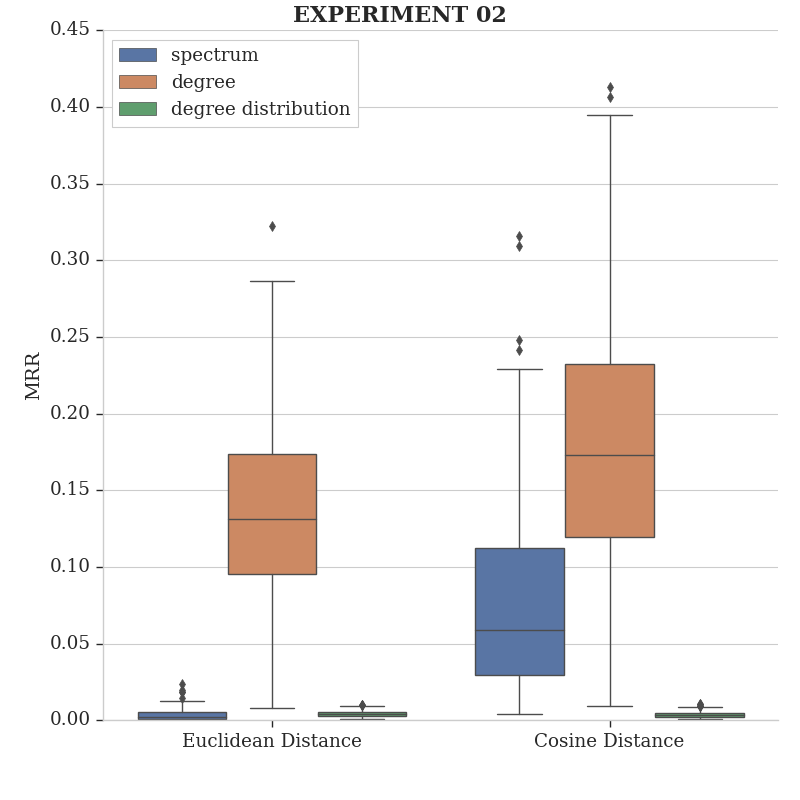}
	\end{center}
	\caption{Mean-reciprocal-rank (MRR) of all mixtures in experiment 02: dataset Dev DSD100, vocal stems and their correspondent mixtures. Pair-wise similarity between the clean vocals and the mixture signals in the magnitude, degree and degree distribution domain for each track. The clean vocal time frames act as query and the expected closest neighbour is that time frame in the mixture.   }
	\label{fig:exp02}
	\vspace{-0.3cm}
\end{figure}
%------------------------------------------------------------------------------------------------------------------

In the second experiment we use the publicly available Demixing Secrets Dataset (DSD100), containing the stems and mixtures of 100 songs sampled at 44100 Hz \cite{liutkus20172016}. In this case the query will be clean vocal frames (i.e. lead harmonic source) and the goal is to find their corresponding frames in the mixture. The magnitude spectrogram for both the vocal and mixture tracks is calculated, with a window size of 2046 samples with 50\% overlap, and only the first 500 frequency bins will be considered in the following (i.e. low-pass filter cut-off at around 10kHz). Based on the spectrogram energy of the vocal stem, we select the frames with vocal activity and use them as query frames. The pair-wise distance between the clean vocal query frames and all the frames in the mixture spectogram is then calculated and sorted. The rank of the corresponding mixture frame containing the clean vocal query is then processed and stored to calculate the MRR. This procedure is repeated for the spectral visibility graph degree representation as well as for the degree distribution.

Figure \ref{fig:exp02} shows the results for experiment 02. The proposed representation is, in both cases (Euclidean and cosine distance), visibly much more suitable than the magnitude spectrogram and the degree distribution for the given task.

 The fact that the degree distribution representation always achieved the worst results shows that the location of the harmonic peaks is a crucial piece of information for this type of harmonic similarity task. 
Even though the degree distribution was not advantageous in this case, there may be other audio analysis tasks for which it is useful, such as those requiring pitch shift-invariance \cite{seetharaman2017cover,FanoYelaEOS18_ShiftInvariant_ICASSP}.

\section{Conclusion}
Here we introduced the visibility graph for magnitude spectra. 
We propose to use the spectral visibility graph degree as an alternative representation for magnitude spectra. 
Such representation presents properties valuable in audio analysis.
Here we focus on a translation invariance of the proposed representation as it directly relates to a harmonic event in presence of broadband noise.
We further demonstrate its use for robust similarity measures of both synthetic and real harmonic events.
Even though we have demonstrated one application of the proposed representation, we expect such graph-based approach for audio analysis to find other useful applications in the future.

\section*{Acknowledgment}
This work was funded by EPSRC grant EP/L019981/1.
Dan Stowell was supported by EPSRC Early Career research fellowship EP/L020505/1.



%\bibliographystyle{IEEEtran}


% \end{thebibliography}

\bibliographystyle{abbrv}
\bibliography{referencesMusic}

\begin{thebibliography}{10}

\bibitem{bang2008digraphs}
J.~Bang-Jensen and G.~Z. Gutin.
\newblock {\em Digraphs: theory, algorithms and applications}.
\newblock Springer Science \& Business Media, 2008.

\bibitem{BarberC2010_Graphicalmodelstime_ISPM}
D.~Barber and A.~T. Cemgil.
\newblock Graphical models for time-series.
\newblock {\em IEEE Signal Processing Magazine}, 27(6):18--28, 2010.

\bibitem{bello2005robust}
J.~P. Bello and J.~Pickens.
\newblock A robust mid-level representation for harmonic content in music
  signals.
\newblock In {\em ISMIR}, volume~5, pages 304--311. Citeseer, 2005.

\bibitem{CampanharoSMRA11_dualityTS_Pone}
A.~S. Campanharo, M.~I. Sirer, R.~D. Malmgren, F.~M. Ramos, and L.~A.~N.
  Amaral.
\newblock Duality between time series and networks.
\newblock {\em PloS one}, 6(8):e23378, 2011.

\bibitem{FanoYelaEFS17_HybridKamNmf_ICASSP}
D.~Fano~Yela, S.~Ewert, D.~FitzGerald, and M.~B. Sandler.
\newblock Interference reduction in music recordings combining kernel additive
  modelling and non-negative matrix factorization.
\newblock In {\em Proceedings of the {IEEE} International Conference on
  Acoustics, Speech, and Signal Processing ({ICASSP})}, New Orleans, USA, 2017.

\bibitem{FanoYelaEOS18_ShiftInvariant_ICASSP}
D.~Fano~Yela, S.~Ewert, K.~O'Hanlon, and M.~B. Sandler.
\newblock Shift-invariant kernel additive modelling for audio source
  separation.
\newblock In {\em Proceedings of the {IEEE} International Conference on
  Acoustics, Speech, and Signal Processing ({ICASSP})}, Calgary, Canada, 2018.

\bibitem{dFanoYela_LVA}
D.~Fano~Yela, D.~Stowell, and M.~Sandler.
\newblock Does k matter? k-nn hubness analysis for kernel additive modelling
  vocal separation.
\newblock In Y.~Deville, S.~Gannot, R.~Mason, M.~D. Plumbley, and D.~Ward,
  editors, {\em Latent Variable Analysis and Signal Separation}, pages
  280--289, Cham, 2018. Springer International Publishing.

\bibitem{LacasaLBLN08_VisibilityGraph_PNAS}
L.~Lacasa, B.~Luque, F.~Ballesteros, J.~Luque, and J.~C. Nuno.
\newblock From time series to complex networks: The visibility graph.
\newblock {\em Proceedings of the National Academy of Sciences},
  105(13):4972--4975, 2008.

\bibitem{LacasaNL15_multiplex_Nature}
L.~Lacasa, V.~Nicosia, and V.~Latora.
\newblock Network structure of multivariate time series.
\newblock {\em Scientific reports}, 5:15508, 2015.

\bibitem{Lan15_DC_Caos}
X.~Lan, H.~Mo, S.~Chen, Q.~Liu, and Y.~Deng.
\newblock Fast transformation from time series to visibility graphs.
\newblock {\em Chaos: An Interdisciplinary Journal of Nonlinear Science},
  25(8):083105, 2015.

\bibitem{liutkus20172016}
A.~Liutkus, F.-R. St{\"o}ter, Z.~Rafii, D.~Kitamura, B.~Rivet, N.~Ito, N.~Ono,
  and J.~Fontecave.
\newblock The 2016 signal separation evaluation campaign.
\newblock In {\em International Conference on Latent Variable Analysis and
  Signal Separation}, pages 323--332. Springer, 2017.

\bibitem{LuqueLBFL09_HVg_PRE}
B.~Luque, L.~Lacasa, F.~Ballesteros, and J.~Luque.
\newblock Horizontal visibility graphs: Exact results for random time series.
\newblock {\em Physical Review E}, 80(4):046103, 2009.

\bibitem{mcfee2014analyzing}
B.~McFee and D.~Ellis.
\newblock Analyzing song structure with spectral clustering.
\newblock In {\em ISMIR}, pages 405--410, 2014.

\bibitem{musmeci2017multiplex}
N.~Musmeci, V.~Nicosia, T.~Aste, T.~Di~Matteo, and V.~Latora.
\newblock The multiplex dependency structure of financial markets.
\newblock {\em Complexity}, 2017, 2017.

\bibitem{NunezLVGL12_PeriodicityHVg_IJBC}
A.~Nu{\~n}ez, L.~Lacasa, E.~Valero, J.~P. G{\'o}mez, and B.~Luque.
\newblock Detecting series periodicity with horizontal visibility graphs.
\newblock {\em International Journal of Bifurcation and Chaos}, 22(07):1250160,
  2012.

\bibitem{NunezLGL12_ReviewVg_InTech}
A.~M. Nu{\~n}ez, L.~Lacasa, J.~P. Gomez, and B.~Luque.
\newblock Visibility algorithms: A short review.
\newblock In {\em New Frontiers in Graph Theory}. InTech, 2012.

\bibitem{seetharaman2017cover}
P.~Seetharaman and Z.~Rafii.
\newblock Cover song identification with 2d fourier transform sequences.
\newblock In {\em 2017 IEEE International Conference on Acoustics, Speech and
  Signal Processing (ICASSP)}, pages 616--620. IEEE, 2017.

\bibitem{sonnleitner2014quad}
R.~Sonnleitner and G.~Widmer.
\newblock Quad-based audio fingerprinting robust to time and frequency scaling.
\newblock In {\em DAFx}, pages 173--180. Citeseer, 2014.

\bibitem{StowellP2013_Segregatingeventstreams_TJoMLR}
D.~Stowell and M.~D. Plumbley.
\newblock Segregating event streams and noise with a markov renewal process
  model.
\newblock {\em The Journal of Machine Learning Research}, 14(1):2213--2238,
  2013.

\bibitem{wang2003industrial}
A.~Wang et~al.
\newblock An industrial strength audio search algorithm.
\newblock In {\em Ismir}, volume 2003, pages 7--13. Washington, DC, 2003.

\end{thebibliography}

\end{document}